\documentclass[conference]{IEEEtran} 
\usepackage[numbers,sort&compress]{natbib}
\usepackage{blindtext}
\usepackage[bottom]{footmisc}
\usepackage{graphicx}
\usepackage{epstopdf}
\usepackage{algorithm}  
\usepackage{algorithmic} 
\usepackage{subfigure}
\usepackage{amsmath,bm}
\usepackage{amssymb}
\usepackage{amsthm}
\usepackage{cases}
\usepackage{setspace}
\usepackage{multirow}

\IEEEoverridecommandlockouts
\begin{document}
	
\title{Resource Allocation for Cooperative D2D-Enabled Wireless Caching Networks}

\author{\IEEEauthorblockN{Jiaqi Liu\IEEEauthorrefmark{1},
		Shengjie Guo\IEEEauthorrefmark{2},
		Sa Xiao\IEEEauthorrefmark{1},
		Miao Pan\IEEEauthorrefmark{3},
		Xiangwei Zhou\IEEEauthorrefmark{2},
		Geoffrey Ye Li\IEEEauthorrefmark{4},
		Gang Wu\IEEEauthorrefmark{1},
		and Shaoqian Li\IEEEauthorrefmark{1}
        }
	
	\IEEEauthorblockA{
		\IEEEauthorrefmark{1}
		National Key Laboratory of Science and Technology on Communications, UESTC, Chengdu, China \\
		\IEEEauthorrefmark{2}
		Division of Electrical and Computer Engineering, Louisiana State University, Baton Rouge, LA, USA \\
		\IEEEauthorrefmark{3}
		Department of Electrical and Computer Engineering, University of Houston, Houston, TX 77204\\
		\IEEEauthorrefmark{4}
		School of Electrical and Computer Engineering, Georgia Institute of Technology, Atlanta, GA, USA
	}
    
}

\maketitle

\begin{abstract}
In this paper, we study the resource allocation problem for a cooperative \textit{device-to-device} (D2D)-enabled wireless caching network,
where each user randomly caches popular contents to its memory  
and shares the contents with nearby users through D2D links.
To enhance the throughput of spectrum-sharing D2D links, 
which may be severely limited by the interference among D2D links, 
we enable the cooperation among some of the D2D links to eliminate the interference among them.
We formulate a joint link scheduling and power allocation problem to maximize the overall throughput of
\textit{cooperative D2D links} (CDLs) and \textit{non-cooperative D2D links} (NDLs), which is NP-hard. 
To solve the problem, we decompose it into two sub-problems 
that maximize the sum rates of the CDLs and the NDLs, respectively. 
For CDL optimization, we propose a semi-orthogonal-based algorithm for joint user scheduling and power allocation. 
For NDL optimization, we propose a novel low-complexity algorithm to perform link scheduling 
and develop a \textit{Difference of Convex functions} (D.C.) programming method to solve the non-convex power allocation problem.
Simulation results show that the cooperative transmission can significantly increase both the number of served users
and the overall system throughput.
\end{abstract}

\section{Introduction}

Due to the exponential growth of mobile devices and mobile services,
the traditional infrastructure of cellular networks cannot fully accommodate 
the high data-rate demands of users.
To relieve the traffic load of core networks as well as to improve the spectrum efficiency of the whole system,
both \textit{device-to-device} (D2D) communications and wireless caching
have been considered as promising techniques in the next generation cellular networks \cite{d2d_feng,EdgeCaching}.

With D2D communications, popular contents cached in a regular mobile device 
can be easily obtained by the desired users nearby,
which significantly improves the network throughput and greatly relieves the traffic pressure on backhaul networks
\cite{EdgeCaching, Molisch_clusterD2D,Molisch_d2d}.
In practice, multiple D2D links in a hotspot area can coexist and share the same time/frequency resources 
due to the short distances of D2D links and low transmit power of mobile devices \cite{spectrum_sharing}.
However, link scheduling and power allocation strategies should be well designed 
to mitigate the inter-link interference that
severely damages the performance of spectrum sharing D2D networks. 
In \cite{Molisch_clusterD2D}, the cell area is divide into clusters and only one D2D link within each cluster is active
to mitigate the inter-link interference.
In \cite{Molisch_d2d}, the D2D users are divide into clusters
and only one cluster from a certain number of adjacent clusters is active
at each time slot to mitigate the inter-cluster interference.
In \cite{lzhang_d2d_caching}, the interference among D2D links are mitigated by
efficient link scheduling and power control strategies.

Cooperative transmission, where multiple transmission nodes serve multiple users together with cooperation,
is adopted in cellular networks to effectively mitigate the inter-link interference \cite{bs_coop,reviewer_add}.
However, cooperative transmission in D2D networks is rarely studied 
because it is usually not feasible for two or more D2D transmitters to have the massage that a D2D receiver requests at the same time
in conventional D2D networks.
In D2D-enabled wireless caching networks, the cooperation among D2D transmitters can be enabled to improve the system throughput 
by utilizing the redundancy of caching, i.e.,  two or more D2D transmitters may cache the same contents \cite{CYang_CoD2D}.
In \cite{CYang_CoD2D}, the authors propose an opportunistic cooperation strategy 
that enables interference-free cooperative D2D communications  all clusters,
given that a certain group of files are cached and requested by users from every cluster.

In this paper, we study the resource allocation problem in
a cooperative D2D-enable wireless caching network.
To maximize the overall throughput of {cooperative D2D links} (CDLs) and \textit{non-cooperative D2D links} (NDLs),
we formulate a joint link scheduling and power allocation problem.
To solve this NP-hard problem, we decompose it into two sub-problems
that maximize the sum rates of the CDLs and the NDLs, respectively. 
For CDL, we propose a semi-orthogonal-based algorithm for joint user scheduling and power allocation. 
For NDL, we propose a novel low-complexity algorithm to perform link scheduling 
and a \textit{Difference of Convex functions} (D.C.) programming method to solve the non-convex power allocation problem.

The remainder of this paper is organized as follows.
Section II describes the system model and the cooperation strategy. 
Section III formulates the optimization problem that maximizes the overall system throughput.
The optimization problem is decomposed into two sub-problems, 
which are solved in Sections IV and V, respectively.
Section VI provides the simulation results.
Section VII concludes this paper.

\section{System Model and Cooperation Strategy}

\subsection{System Model}

\begin{figure}
	\centering
	\includegraphics[scale=0.5]{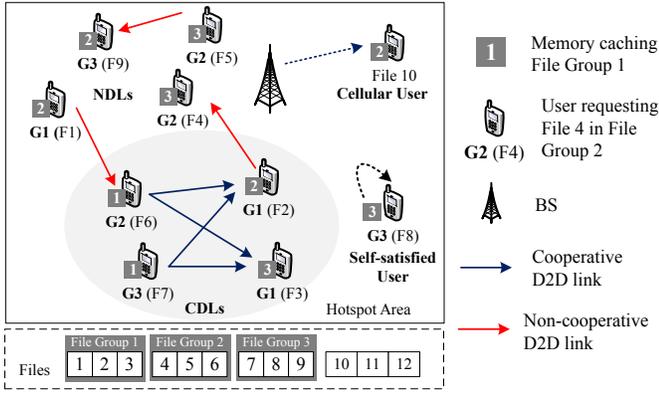}
	\caption{System model for a cooperative D2D-enabled caching network,
		where $K=10$, $N_F=12$, $N_0 = 3$, and $G=3$.}
	\label{system_model}
\end{figure}

Suppose that $K$ single-antenna users $\mathcal{K} = \{1,\cdots,K\}$
are randomly distributed in a hotspot of a cell as shown in Fig. \ref{system_model}.
There are $N_F$ files in the system.
Each user has a memory with uniformed size to cache $N_0$ files and can share its cached files to nearby users via D2D communications.
The users request the files according to Zipf distribution with parameter $\beta$,
which indicates that the $\eta$th file is requested by each user with probability
${\eta^{-\beta}}\Big/{\sum_{\theta=1}^{N_F}\theta^{-\beta}}, \forall \eta = 1,\cdots,N_F$.
According to the memory size of each user, we divide the most popular $N_\text{popular}$ files into $G = N_\text{popular} / N_0$ groups, 
where the $g$th file group $\mathcal{G}_g$ contains the $(g-1)N_0+1$th to the $gN_0$th files.
Then the probability that a user requests a file within the $g$th file group $\mathcal{G}_g$ is \cite{CYang_CoD2D}
\begin{align}
P^\text{r}_g = \frac{\sum_{\eta = (g-1)N_0+1 }^{gN_0} \eta^{-\beta}}{\sum_{\theta=1}^{N_F} \theta^{-\beta}}.
\end{align}

We define a file group request matrix $\mathbf{Y}$, where entry
$y_{k,g}=1$ if a file in $\mathcal{G}_g$ is requested by user $k\in\mathcal{K}$, and $y_{k,g}=0$ otherwise.
We denote $g^\text{r}(k)$ as the file group requested by the $k$th user.

Similar to \cite{CYang_CoD2D}, we assumes that
each user caches each file group with uniform probability $1/K$.
We define a file group caching matrix $\mathbf{X}$, where the entry
$x_{k,g}=1$ if user $k\in\mathcal{K}$ caches $\mathcal{G}_g$, and $x_{k,g}=0$ otherwise.

Based on $\mathbf{X}$ and $\mathbf{Y}$, we define
$\mathcal{M}_g = \{ k\in\mathcal{K} | x_{k,g} = 1 \}$ 
as the set of the users that cache $\mathcal{G}_g$
and $\mathcal{N}_g = \{ k\in\mathcal{K} | y_{k,g} = 1, x_{k,g} = 0 \}$
as the set of the users that request the files in $\mathcal{G}_g$
but do not cache $\mathcal{G}_g$.

\subsection{Cooperative Content Delivery Policy}

Based on the source where a user can fetch its requested file,
we classify the users into three categories: 
\textit{self-satisfied users}
	who request the files that are cached by themselves
	and can obtain them directly from their own memory;
\textit{D2D users}
	who request the files that are not cached by themselves
	but cached by their nearby users
	and can obtain the files via D2D communications;
and \textit{cellular users}
	who request the files that are neither cached by themselves nor cached by their nearby users
	and have to request the files from the BS.
In this paper, we only focus on the D2D users
who can fetch their requested files via either \textit{cooperative D2D links} (CDLs) or
\textit{non-cooperative D2D links} (NDLs).

We designate one file group to be delivered via CDLs
and the other file groups to be delivered via NDLs \cite{CYang_CoD2D}.
This is because the improved performance of cooperation 
is evident with enough numbers of transmitters and receivers.
Caching systems generally consider the scenario where
the first few popular files account for the majority of requests \cite{EdgeCaching}.
Therefore, it is usually not worthy to enable the cooperative transmission of another file group
that is requested by very few users.
On the other hand, our proposed link scheduling and power allocation algorithms
are also applicable when cooperation of multiple file groups is enabled.
We denote the transmission mode indicator $t_g$ for each $\mathcal{G}_g$, 
where $t_g = 1$ if $\mathcal{G}_g$ is transmitted via CDLs, and $t_g = 0$ otherwise.

To mitigate the interference between CDLs and NDLs,
two separate frequency bands are used for CDLs and NDLs independently.
As mentioned before, there is no interference among CDLs.
However, there is interference among NDLs, which will be mitigated 
with our proposed power allocation and link scheduling methods.
We assume that the frequency bands allocated for CDLs and NDLs are preset and fixed.
Dynamic frequency allocation is another interesting topic but not the focus of this paper.

\section{Problem Formulation and Analysis}

In this section, we first derive the achievable rates of CDLs and NDLs, respectively,
and then formulate an optimization problem to maximize the overall system throughput.

\subsection{Achievable Rates of CDLs}

Let $\mathcal{T}^\text{C}$ and $\mathcal{R}^\text{C}$ denote the sets of
\textit{cooperative D2D transmitters} (CTs) and \textit{cooperative D2D receivers} (CRs), respectively.
The achievable rate of CR $n$ is
\begin{align}\label{coop_rate}
	R_n^\text{C} =  W_\text{C} \log_2 \left( 1+
	\frac{ P_n |\mathbf{h}_n^\text{H} \mathbf{\bar{w}}_n|}
	{\sum_{k\in\mathcal{R}^\text{C}/n} P_k|\mathbf{h}_n^\text{H}\mathbf{\bar{w}}_k|+N^\text{C}_n}\right),
\end{align}
where $W_\text{C}$ denotes the bandwidth allocated for CDLs,
$P_n$ denote the total transmit power allocated to CR $n$ by all CTs,
$\mathbf{h}_n\in\mathbb{C}^{|\mathcal{T}^\text{C}|}$ denotes the channel vector from the CTs to CR $n$,
$\mathbf{\bar{w}}_n\in\mathbb{C}^{|\mathcal{T}^\text{C}|}$ denotes the normalized precoding vector of CR $n$,
and $N^\text{C}_n$ denotes the noise power at CR $n$.

We adopt \textit{zero-forcing beamforming} (ZFBF) \cite{sus_Goldsmith} such that
$\mathbf{W} = \{\mathbf{w}_n\}_{n\in\mathcal{R}^\text{C}} = \mathbf{H}\left( \mathbf{H}^\text{H}\mathbf{H}\right)^{-1} $,
where $\mathbf{H} = \{ \mathbf{h}_n \}_{n\in\mathcal{R}^\text{C}} $,
and $\mathbf{\bar{w}}_n =\frac{\mathbf{w}_n}{\|\mathbf{w}_n\|^2}$.
Let $\bar{w}_{m,n}$ and $h_{m,n}$ denote the $m$-th entries in $\bar{\mathbf{w}}_n$ and $\mathbf{h}_n$, respectively.
Equation (\ref{coop_rate}) becomes
\begin{align}
R^{\text{C}}_n = W_\text{C} \log_2 \left( 1+ \frac1{N^\text{C}_n} 
	\sum_{m\in\mathcal{T}^\text{C}} P_{n}|\bar{w}_{m,n}|^2\left|h_{m,n}\right|^2\right).
\end{align} 

\subsection{Achievable Rates of NDLs}

We assume that an NDL can only be established between two users within distance $r$,
which is referred as D2D radius.
The set of potential \textit{non-cooperative D2D transmitters} (NTs) for user 
$j\in\textstyle\bigcup\nolimits_{f=1}^{F} (1-t_g)\mathcal{N}_g$ is
\begin{equation}
\widehat{\mathcal{T}}^\text{N}_j = \left\lbrace k \in \mathcal{M}_{g^\text{r}(j)} ~\big|~
d(k,j)<r\right\rbrace ,
\end{equation}
where $d(k,j)$ denotes the distance between users $k$ and $j$.

Obviously, user $j$ can be a potential \textit{non-cooperative D2D receiver} (NR) only if 
it has at least one potential NT,
i.e., $\widehat{\mathcal{T}}^\text{N}_j \neq \varnothing$. 
We denote the set of potential NRs as
\begin{equation}
\widehat{\mathcal{R}}^\text{N} = \left\lbrace j \in (1-t_g) \mathcal{N}_g ~\big|~
\widehat{\mathcal{T}}^\text{N}_j \neq \varnothing \right\rbrace.
\end{equation}

Let $\mathcal{T}^\text{N}$ and $\mathcal{R}^\text{N}$ denote the sets of finally selected NTs and NRs, respectively.
Let $\tau(j)\in\widehat{\mathcal{T}}^\text{N}_j$ denote the corresponding NT of NR $j$
and $h(i,j)$ denote the channel coefficient from $\tau(i)$ to NR $j$.
The \textit{signal-to-interference-plus-noise-ratio} (SINR) at NR $j$ is
\begin{equation}
	\gamma_j = \frac{p_j|h(j,j)|^2}
	{\sum_{i\in {\mathcal{R}^\text{N}}/j} p_i|h(i,j)|^2 + N^\text{N}_j},
\end{equation}
where $p_j$ denotes the transmit power of $\tau(j)$
and $N^\text{N}_j$ denotes the noise power at NR $j$.
The achievable rate of NR $j$ is expressed as
$R_j^\text{N} = W_\text{N} \log_2(1+\gamma_j)$,
where $W_\text{N}$ is the bandwidth allocated for NDLs.

\subsection{System Throughput Optimization Problem}

We aim to maximize the system throughput, i.e., the sum rate of all CDLs and NDLs.
The problem is formulated as
\begin{align}\label{orginal_problem}
	& \max\limits_{ \substack{ \mathcal{R}^\text{C},\mathcal{T}^\text{C}, \mathbf{P},\\   
	\mathcal{R}^\text{N},\mathcal{T}^\text{N}, \mathbf{p} }} ~~
	\sum_{m \in \mathcal{R}^\text{C}} R_m^\text{C} 
		+ \sum_{j \in \mathcal{R}^\text{N}} R_j^\text{N}\\
	\label{ConstCoopInd} \tag{\ref{orginal_problem}a} 
	&~~\text{s.t.}~~ t_g\in\{0,1\}, \forall g, \\
	\label{ConstNumCoop} \tag{\ref{orginal_problem}b} 
	&~~~~~~~  \textstyle \sum_{f=1}^{F}  t_g = 1, \\
	\label{ConstUc} \tag{\ref{orginal_problem}c} 
	&~~~~~~~ \mathcal{T}^\text{C} \subseteq \textstyle\bigcup\nolimits_{g=1}^{G} t_g \mathcal{M}_g,
		~ \mathcal{R}^\text{C} \subseteq \textstyle\bigcup\nolimits_{g=1}^{G} t_g \mathcal{N}_g,\\
	\label{ConstUn} \tag{\ref{orginal_problem}d} 
	&~~~~~~~ \mathcal{R}^\text{N} \subseteq \widehat{\mathcal{R}}^\text{N},~ \tau(j)\in \widehat{\mathcal{T}}^\text{N}_j,~\forall j \in \mathcal{R}^\text{N},\\
	\label{ConstHD} \tag{\ref{orginal_problem}e} 
	&~~~~~~~  \mathcal{T}^\text{N} \cap \mathcal{R}^\text{N} = \varnothing,\\
	\label{ConstCoopPwr} \tag{\ref{orginal_problem}f} 
	&~~~~~~~  P_n \geq 0, ~\forall n, ~\textstyle\sum_{n\in\mathcal{R}^\text{C}} P_n |w_{m,n}|^2 \leq p_m^\text{max}, \forall m,\\
	\label{ConstNonCoopPwr} \tag{\ref{orginal_problem}g} 
	&~~~~~~~  0 \leq p_j \leq p_j^\text{max}, \forall j\in\mathcal{R}^\text{N},\\
	\label{ConstCoopQoS} \tag{\ref{orginal_problem}h} 
	&~~~~~~~  R_m^\text{C} \geq R_m^\text{min}, \forall m \in \mathcal{T}^\text{C},\\
	\label{ConstNonCoQoS} \tag{\ref{orginal_problem}i} 
	&~~~~~~~  R_j^\text{N} \geq R_j^\text{min}, \forall j \in \mathcal{T}^\text{N},
\end{align}
where $\mathbf{P}$ and $\mathbf{p}$ are the vectors consisting of all $P_n$'s and ${p}_j$'s,
respectively,
constraint (\ref{ConstNumCoop}) indicates that only one file group is delivered through CDLs,
constraints (\ref{ConstUc}) and (\ref{ConstUn}) ensure the validity of cooperative and non-cooperative D2D users, respectively,
constraint (\ref{ConstHD}) indicates that each NT cannot be an NR at the same time due to half-duplex,
constraints (\ref{ConstCoopPwr}) and (\ref{ConstNonCoopPwr}) are the peak transmit power constraints for CTs and NTs, respectively,
and constraints (\ref{ConstCoopQoS}) and (\ref{ConstNonCoQoS}) are the \textit{Quality-of-Service} (QoS) constraints for CRs and NRs, respectively.

Due to the computational complexity to solve problem (\ref{orginal_problem}),
we divide it into two sub-problems,
which will be solved in Sections \ref{section_cdl} and \ref{section_ndl}, respectively.

\section{CDL Scheduling and Power Allocation}\label{section_cdl}

In this section, we solve the first sub-problem of problem (\ref{orginal_problem}),
which considers CDL optimization as follows:
\begin{align}\label{noncoop_problem}
\max\limits_{ \mathcal{R}^\text{C},\mathcal{T}^\text{C}, \mathbf{P}} & ~ \sum_{n \in \mathcal{R}^\text{C}} R_n^\text{C} \\\nonumber
\text{s.t.}& ~~ \text{(\ref{ConstCoopInd}),~(\ref{ConstNumCoop}),
			(\ref{ConstUc}),~(\ref{ConstCoopPwr}),~(\ref{ConstCoopQoS})}.
\end{align}

To solve problem (\ref{noncoop_problem}), we first select the CTs and the CRs
and then solve the power allocation problem to maximize the sum rate of the selected CRs.

\subsection{CT and CR Selection}

Based on the analysis in \citep[Theorem 1]{sus_Goldsmith} and the uniform caching probability,
we heuristically choose the mostly requested file group as cooperatively transmitted file group,
which is
\begin{equation}
	g^* = \max_{1 \leq g \leq G} |\mathcal{N}_g|.
\end{equation}

\begin{algorithm}[!tp]   
	\caption{Semi-orthogonal CDL Scheduling Algorithm}   
	\label{alg_coop_sus}   
	\begin{algorithmic}[1]  
		\REQUIRE Channel vectors $\mathbf{h}_n, \forall n\in\mathcal{N}_{g^*}$.
		\ENSURE Selected cooperative D2D receivers $\mathcal{R}^\text{C}$.
		\STATE \textit{\textbf{Initialize:}}
		$\Omega_1 = \mathcal{N}_{g^*};~ i = 1;~ \mathcal{R}^\text{C} = \varnothing;$
		\WHILE {$i < |\mathcal{M}_{g^*}|$ \AND $\Omega_i\neq \varnothing$}
		\FOR {$t \in \Omega_i$}
		\IF {$i=1$}
		\STATE $\mathbf{g}_t = \mathbf{h}_t$;
		\ELSE
		\STATE $\mathbf{g}_t = \mathbf{h}_t - \sum_{j=1}^{i-1}
		\frac{\mathbf{\tilde{g}}_{(j)}^\text{H} \mathbf{h}_t}
		{\|\mathbf{\tilde{g}}_{(j)}\|^2} \mathbf{\tilde{g}}_{(j)}$;$~^1$
		\ENDIF
		\ENDFOR
		\STATE $\pi(i) = \arg\max\limits_{t\in\Omega_i} \|\mathbf{g}_t\|^2$;\
		$\mathcal{R}^\text{C}\leftarrow \mathcal{R}^\text{C}\cup \{\pi(i)\}$;
		\STATE Solve the power allocation problem (\ref{subprob_pa});
		\IF {problem (\ref{subprob_pa}) is not feasible}
		\STATE $\mathcal{R}^\text{C}\leftarrow \mathcal{R}^\text{C}  \big/ \pi(i)$;
		\STATE break;
		\ENDIF
		\STATE  $\mathbf{\tilde{g}}_{(i)}=\mathbf{g}_{\pi(i)}$; 
		\STATE $\Omega_{i+1} = \big\lbrace t \in \Omega_{i} \big/ \pi(i) ~\big|~
		\frac{\left| \mathbf{h}^\text{H}_t\mathbf{\tilde{g}}_{(i)} \right|}
		{\|\mathbf{h}_t\| \|\mathbf{\tilde{g}}_{(i)}\|}
		< \epsilon \big\rbrace $;
		\STATE $i \leftarrow i+1$;  
		\ENDWHILE
	\end{algorithmic}  
\end{algorithm}
\footnotetext{$^1$ The first run of Line 10 happens within the \textit{while loop} of $i=2$,
	before which the value of $\mathbf{\tilde{g}}_{(1)}$ has been initialized
	within the previous \textit{while loop} of $i=1$.}

We let all users in $\mathcal{M}_{g^*}$ be the CTs, i.e., $\mathcal{T}^\text{C} = \mathcal{M}_{g^*}$.
Inspired by \textit{semi-orthogonal user selection} (SUS) in 
\textit{multi-user multiple-input-multiple-output} (MU-MIMO) systems \cite{sus_Goldsmith},
we develop a semi-orthogonal CR scheduling algorithm to iteratively select $\mathcal{R}^\text{C}$ from $\mathcal{N}_{g^*}$, 
which is summarized as \textbf{Algorithm \ref{alg_coop_sus}}.
We select one CR and optimize the power allocation over the selected CRs in each iteration.
The iteration terminates once the number of CRs reaches the number of CTs,
the set of unselected CRs is empty or the power allocation problem is not feasible.

\subsection{Power Allocation for Scheduled CRs}

We solve the power allocation problem that 
maximizes the sum rate of the selected CRs, which is
\begin{align}\label{subprob_pa}
\max_{\mathbf{P}} 
&~~ \textstyle\sum_{n\in\mathcal{R}^\text{C}} R^{\text{C}}_n
\\\nonumber
\text{s.t.} &~~ \text{(\ref{ConstCoopPwr}),~(\ref{ConstCoopQoS})}.
\end{align}

Using the Lagrangian decomposition method \cite{boyd2004convex}, 
we obtain the Lagrangian function of problem (\ref{subprob_pa}), 
which is
\begin{align}\label{lag_fun}
L(\mathbf{p}^\text{C}, \mathbf{\lambda}, \mathbf{\mu}) = 
& \textstyle\sum_{n\in\mathcal{R}^\text{C}} R^{\text{C}}_n
\nonumber \\
&- \textstyle\sum_{m\in\mathcal{T}^\text{C}} \textstyle\lambda_m \left( \sum_{n\in\mathcal{R}^\text{C}} P_n|w_{m,n}|^2 - p_m^\text{max} \right) 
\nonumber \\
& +\textstyle\sum_{n\in\mathcal{R}^\text{C}} \mu_n \left( R^{\text{C}}_n - R_\text{min}\right),
\end{align}
where $\lambda_m$ and $\mu_n$ are the introduced Lagrange multipliers.

According to the KKT conditions, 
the optimal power allocation of problem (\ref{subprob_pa}) is
\begin{align}
	P_{n}^* = \min \{ P_0, P_n^\text{max} \},
\end{align}
where
\begin{align}
	P_0 = \left[ \frac{1+\mu_n}{\ln 2 \sum_{m\in\mathcal{T}^\text{C}} \lambda_m |\bar{w}_{m,n}|^2}
			-\frac{N^\text{C}_n}{\sum_{m\in\mathcal{T}^\text{C}} |\bar{w}_{m,n}|^2|h_{m,n}|^2} 
		\right]^+
\end{align}
and $[x]^+ = \max\{0,x\}$.

We solve the multipliers $\lambda_m$'s and $\mu_n$'s iteratively using gradient descent,
where the multipliers in the $t$th iteration are updated as follows:
\begin{gather}
	\lambda^{(t+1)}_m = \left[ \lambda^{(t)}_m 
		- \varsigma^{(t)} \left( P_n^\text{max} - \textstyle\sum_{n\in\mathcal{R}^\text{C}} P_n|\bar{w}_{m,n}|^2\right) \right]^+,\\
	\mu^{(t+1)}_n = \left[ \mu^{(t)}_n 
		- \varrho^{(t)}\left(R_n^\text{C}-R_\text{min}\right)\right]^+,
\end{gather}
where $\varsigma^{(t)}$ and $\varrho^{(t)}$ are small positive step sizes for step $t$.

\section{NDL Scheduling and Power Allocation}\label{section_ndl}

After the CDLs are established,
we establish NDLs among the remaining users.
The optimization problem is given by
\begin{align} \label{problem_Nc}
	\max\limits_{ \mathcal{T}^\text{N},\mathcal{R}^\text{N}, \mathbf{p} } ~ 
		&  \sum_{j \in \mathcal{R}^\text{N}} R_j^\text{N}, 
    \\ \label{qos_gamma} \tag{\ref{problem_Nc}a}
		\text{s.t.}~~ &~ \gamma_j \geq \bar{\gamma}_j, ~\forall j,
    \\\nonumber
    	&~ \text{(\ref{ConstUn}),~(\ref{ConstHD}),
			(\ref{ConstNonCoopPwr})},
\end{align}
where (\ref{qos_gamma}) is the minimum SINR constraint transformed from the QoS constraint (\ref{ConstNonCoQoS})
and $\bar{\gamma}_j = 2^{R_j^\text{min}}-1$.

To solve problem (\ref{problem_Nc}), 
we first schedule as many NDLs satisfying the constraints as possible  
and then perform power allocation to maximize the minimum rate of the NDLs.

\subsection{NDL Scheduling}

The NDL scheduling sub-problem of problem (\ref{problem_Nc}) is 
\begin{align}\label{prob_ls}
	\max\limits_{\mathcal{R}^\text{N} \subseteq \widehat{\mathcal{R}}^\text{N}} ~ 
		& \left| \mathcal{R}^\text{N} \right| 
\\ \nonumber
	\text{s.t.}&~ \text{(\ref{ConstHD}),~(\ref{ConstNonCoopPwr}),~(\ref{qos_gamma})},
\end{align}
which can be regarded as an admission control problem.
In regular admission control problems, 
each node is either a potential transmitter or a potential receiver.
However, in our system model, there is a chance that
a certain user can be both a potential NT and a potential NR.
We refer such a user as an ambiguous user.
Since any two scheduled NDLs cannot share the same user due to half duplex,
the problem becomes more complicated.
We propose a novel low-complexity NDL scheduling algorithm to solve the problem,
which is described as follows. 

First, we decide whether each ambiguous user should be an NT or NR.
After this decision is done among all ambiguous users,
the original admission control problem is simplified as a regular admission control problem.
Then we establish as many potential NDLs as possible
and check whether the QoS constraints of the potential NDLs can be satisfied with certain power allocation.
If the QoS constraints of the potential NDLs cannot be satisfied with any power allocation, 
we iteratively remove some NDLs until the QoS constraints of every potential NDL can be satisfied.
The detailed procedure of NDL scheduling algorithm is summarized in \textbf{Algorithm \ref{ndl_schedule}}
and described in the remainder of this subsection.

\begin{algorithm}[tp]   
	\caption{NDL Scheduling Algorithm}   
	\label{ndl_schedule}   
	\begin{algorithmic}[1]  
		\REQUIRE $\widehat{\mathcal{R}}^\text{N},~ \widehat{\mathcal{T}}^\text{N}_j,\forall j\in \widehat{\mathcal{R}}^\text{N}$ and channel gains of all NDLs. \\
		\ENSURE $\mathcal{R}^\text{N}$ and $\mathcal{T}^\text{N}$.\\
		\textbf{\textit{-- Phase I: NT-NRT Decision}}
		\FOR {each $u$}
		\STATE Compute $\alpha_u$ and $\beta_u$ according to equations (\ref{cost_tx}) and (\ref{cost_rx}), respectively;
		\IF {$\alpha_u > \beta_u$}
		\STATE $\widehat{\mathcal{T}}^\text{N}_j \leftarrow \widehat{\mathcal{T}}^\text{N}_j \big/ u,~\forall j \in \widehat{\mathcal{R}}^\text{N}$;
		\ELSE
		\STATE $\widehat{\mathcal{R}}^\text{N} \leftarrow \widehat{\mathcal{R}}^\text{N} \big/ u$;
		\ENDIF
		\ENDFOR 
		\\ \textbf{\textit{-- Phase II: Link Selection}}
		\STATE Form the bipartite graph $G$ consisting of $\widehat{\mathcal{R}}^\text{N}$, $\widehat{\mathcal{T}}^\text{N}$ and potential NDLs;
		\STATE Find the sub-graph $G'$ consisting of the vertices with degrees over 1;
		\STATE Add NTs and NRs in $G-G'$ into $\mathcal{T}^\text{N}$ and $\mathcal{R}^\text{N}$, respectively;
		\STATE Execute \textit{maximum weighted matching} on $G'$, and 
			add the matched NTs and NRs into $\mathcal{T}^\text{N}$ and $\mathcal{R}^\text{N}$, respectively;
		\STATE Compute $\mathbf{p}'$ according to equation (\ref{theoretical_power}).
		\\ \textbf{\textit{-- Phase III: Link Removal}}
		\WHILE {$\mathbf{0}\preceq {\mathbf{p}'} \preceq \bar{\mathbf{p}}$ does not hold}
			\STATE Compute $\xi_u$ and $\zeta_u$ for each user $u\in \mathcal{T}^\text{N}$
				according to equations (\ref{eq_xi}) and (\ref{eq_zeta}), respectively;
			\STATE Find NR $u^*$ to be removed according to equation (\ref{removed_user});
			\STATE $\mathcal{R}^\text{N}\leftarrow \mathcal{R}^\text{N} \big/ u^*$,
					$\mathcal{T}^\text{N}\leftarrow \mathcal{T}^\text{N} \big/ \tau(u^*)$;
			\STATE Update $\mathbf{p}'$ according to equation (\ref{theoretical_power});
		\ENDWHILE
	\end{algorithmic} 
\end{algorithm}

\subsubsection{NT-NR Decision}

For ambiguous user $u$, we calculate the minimum interference levels it introduces to the entire system
when it is selected as an NT and an NR, respectively,
and make a decision by comparing those two interference levels.

Assume that user $u$ is an NT and transmits the file to user $v$
who has the largest channel gain to user $u$
among the users requesting the files that user $u$ caches.
To satisfy (\ref{qos_gamma}),
the transmit power of user $u$ must be at least
$\frac{N_v \bar{\gamma}_v}{\left| h(v,v) \right|^2 }$. 
Therefore, when user $u$ is an NT, the total interference that it introduces to the users in $\widehat{\mathcal{R}}^\text{N}$ is at least 
\begin{equation}\label{cost_tx}
\alpha_{u} 
			= \frac{N_v\bar{\gamma}_u}{\left| h(v,v) \right|^2 }
				\sum_{w \in \widehat{\mathcal{R}}^\text{N}} \left| h(v,w) \right|^2.
\end{equation}

Assume that user $u$ is an NR and
$\tau(u)$ is selected as the user with the largest channel gain to user $u$ among the users caching $g^\text{r}(u)$.
To satisfy (\ref{qos_gamma}),
the transmit power of $\tau(u)$ must be at least 
$\frac{N_u \bar{\gamma}_u}{\left| h(u,u) \right|^2 }$. 
Therefore, when user $u$ is an NR, the total interference that $\tau(u)$ introduces to the users in $\widehat{\mathcal{R}}^\text{N}$ is at least 
\begin{equation}\label{cost_rx}
\beta_u 
		= \frac{N_u\bar{\gamma}_u}{\left| h(u,u) \right|^2 }
			\sum_{w \in \widehat{\mathcal{R}}^\text{N}} \left| h(u,w) \right|^2.
\end{equation}

Based on $\alpha_u$ and $\beta_u$, we make the following decision:
\begin{itemize}
	\item if $\alpha_u < \beta_u$, user $u$ is selected as a potential NT;
	\item otherwise, user $u$ is selected as a potential NR.
\end{itemize}

\subsubsection{Link Selection}

There may be a case that 
some potential NTs have more than one potential NRs and vice versa.
Therefore, we perform link selection to ensure that each NT combines with at most one NR
and vice versa.

We denote a bipartite graph $G(\widehat{\mathcal{T}}^\text{N}, \widehat{\mathcal{R}}^\text{N}, \mathcal{L})$,
where $\widehat{\mathcal{T}}^\text{N} = \bigcup_{j\in\widehat{\mathcal{R}}^\text{N}} \widehat{\mathcal{T}}^\text{N}_j$ and
$\mathcal{L}$ is the set of potential NDLs between $\widehat{\mathcal{T}}^\text{N}$ and $\widehat{\mathcal{R}}^\text{N}$.
$\widehat{\mathcal{T}}^\text{N}$ and $\widehat{\mathcal{R}}^\text{N}$ are two disjoint and independent vertex sets of $G$,
and $\mathcal{L}$ is the edge set of $G$.

Note that there are two types of vertices in $G$:
the vertices with the degree 1 and the vertices with degrees larger than 1.
We assume that the vertices with degrees larger than 1 together with their connected edges
form a sub-graph $G'$.
Obviously, the users in $G-G'$ are involved in only one NDL
and the users in $G'$ are involved in more than one NDLs.
To ensure that each user is involved in at most one NDL,
we select the NDLs in $G'$ with maximum weighted matching algorithm \cite{max_matching},
where the weight of each edge in $G'$ is defined as 
the reciprocal of the channel gain of each corresponding potential NDL.

\subsubsection{Link Checking}
Suppose that $N$ potential NDLs are established between 
$N$ potential NTs and $N$ potential NRs satisfying (\ref{ConstHD}).
If (\ref{qos_gamma}) can be satisfied with the selected NDLs,
there must be power allocation $\mathbf{p}'= [p_{k_1}, \cdots, p_{k_N}]^\text{T}$
satisfying $H(\mathbf{\Gamma})\mathbf{p}' = \mathbf{N}$, where 
$\mathbf{\Gamma} = [\bar{\gamma}_{k_1}, \cdots, \bar{\gamma}_{k_N}]^\text{T}$,
\begin{align}
&H(\mathbf{\Gamma}) = \nonumber\\
&\left[\begin{matrix}
	\frac1{\bar{\gamma}_{k_1}}{|h(k_1,k_1)|^2} &-|h(k_1,k_2)|^2 & \cdots & -|h(k_1,k_N)|^2 \\
	-|h(k_2,k_1)|^2 &\frac1{\bar{\gamma}_{k_2}}{|h(k_2,k_2)|^2} & \cdots & -|h(k_2,k_N)|^2 \\
	\vdots & \vdots &\ddots & \vdots \\
	-|h(k_N,k_1)|^2 &-|h(k_N,k_2)|^2 & \cdots & \frac1{\bar{\gamma}_{k_N}}{|h(k_N,k_N)|^2}
\end{matrix}\right]
\end{align}
and $\mathbf{N}= [N_{k_1}, \cdots, N_{k_N}]^\text{T}$.

To check whether (\ref{ConstNonCoopPwr}) can be satisfied,
we derive 
\begin{equation} \label{theoretical_power}
	\mathbf{p}' = H^{-1}(\mathbf{\Gamma}) \mathbf{N}.
\end{equation}
Let $\bar{\mathbf{p}} =[p^\text{max}_{k_1}, \cdots, p^\text{max}_{k_N}]^\text{T}$
and $\mathbf{0}$ be the $N\times 1$ vector with all zero elements.
We have the following link checking:
\begin{itemize}
\item 
	if $\mathbf{0}\preceq {\mathbf{p}'} \preceq \bar{\mathbf{p}}$,
	constraints (\ref{qos_gamma}) and (\ref{ConstNonCoopPwr}) can be satisfied with the selected potential NDLs;
\item
	if $\mathbf{0}\preceq {\mathbf{p}'} \preceq \bar{\mathbf{p}}$ does not hold,
	constraints (\ref{qos_gamma}) and (\ref{ConstNonCoopPwr}) cannot be satisfied with the selected potential NDLs.
\end{itemize}

\subsubsection{Link Removal}

If $\mathbf{0}\preceq {\mathbf{p}'} \preceq \bar{\mathbf{p}}$
does not hold for $\mathbf{p}'$ in (\ref{theoretical_power}),
we will remove some potential NDLs.
We iteratively remove one potential NDL 
and check whether the remaining potential NDLs are feasible for problem (\ref{prob_ls}) in each iteration.
The link removal procedure is described as follows.

For NR $u$, to satisfy $\bar{\gamma}_{u}$,
the transmit power of $\tau(u)$ must be at least
$\frac{N_u \bar{\gamma}_u}{\left| h(u,u) \right|^2 }$,
which causes at least $\frac{N_u \bar{\gamma}_u}{\left| h(u,u) \right|^2 } \left| h(u,v) \right|^2$ 
amount of interference to user $v\in\mathcal{R}^\text{N} \big/ u$. 
Note that an NR $v$ with lower $\bar{\gamma}_v$ and larger peak transmit power of $\tau(v)$ 
can tolerate more interference.
We define the relative interference from $\tau(u)$ to NR $v$ as
$I_r(u,v) = \frac{ \bar{\gamma}_v }{p_v^\text{max}}
	\frac{N_u \bar{\gamma}_u}{\left| h(u,u) \right|^2 } \left| h(u,v) \right|^2$.
Then the minimum total relative interference caused by $\tau(u)$ is
\begin{equation}\label{eq_xi}
	\xi_u = \sum_{v\in\mathcal{R}^\text{N} / u}  I_r(u,v)
	 =  \frac{N_u \bar{\gamma}_u}{\left| h(u,u) \right|^2 } 
	 \sum_{v\in\mathcal{R}^\text{N} / u}
		 \frac{ \bar{\gamma}_v }{p_v^\text{max}} \left| h(u,v) \right|^2.
\end{equation}

For NR $v\in\mathcal{R}^\text{N} / u$, to satisfy $\bar{\gamma}_{v}$,
the transmit power of $\tau(v)$ must be at least
$\frac{N_v \bar{\gamma}_v}{\left| h(v,v) \right|^2 }$.
Then the minimum total relative interference received by NR $u$ is
\begin{equation}\label{eq_zeta}
\zeta_u =  \frac{ \bar{\gamma}_u }{p_u^\text{max}}
			\sum_{v\in\mathcal{R}^\text{N} / u}
		 \frac{N_v \bar{\gamma}_v}{\left| h(v,v) \right|^2 }  \left| h(v,u) \right|^2.
\end{equation}

Obviously, we should remove NR $u^*$, where
\begin{equation}\label{removed_user}
	u^* = \arg\max\limits_{u\in\mathcal{R}^\text{N}} \max \left\lbrace  
			\xi_u , \zeta_u \right\rbrace,
\end{equation}
which is likely to cause the strongest interference to other NRs 
or receive the strongest interference from other NTs.

\subsection{Power Allocation}

When NDLs are selected, 
problem (\ref{problem_Nc}) is simplified as a power allocation problem,
which is
\begin{align} \label{problem_NcPa}
\max\limits_{ \mathbf{p}} & ~ \sum_{j \in \mathcal{R}^\text{N}} R_j^\text{N} \\ \nonumber
\text{s.t.} & ~~\text{(\ref{ConstNonCoopPwr}),~(\ref{ConstNonCoQoS})}.
\end{align}

However, problem (\ref{problem_NcPa}) tends to allocate more power to the NDLs with good channels
while the achievable rate of the other NDLs may be too low, i.e., the NDLs are not fairly treated. 
Therefore, we modify problem (\ref{problem_NcPa}) based on max-min optimization, which is formulated as
\begin{align} \label{problem_NcPa_mm}
\max\limits_{ \mathbf{p}} & ~  \min\limits_{j \in \mathcal{R}^\text{N}} R_j^\text{N} \\ \nonumber
\text{s.t.}& ~~\text{(\ref{ConstNonCoopPwr})}.
\end{align}

The D.C. programming method \cite{dc_pwralloc}
is adopted to solve problem (\ref{problem_NcPa}), which is transformed to
\begin{align} \label{problem_NcPa_dc}
\max\limits_{\mathbf{p}} & ~ f(\mathbf{p}) -g(\mathbf{p}) \\ \nonumber
\text{s.t.}& ~~\text{(\ref{ConstNonCoopPwr})},
\end{align}
where $f(\mathbf{p}) = \min\limits_{j \in \mathcal{R}^\text{N}} \left\lbrace f_j(\mathbf{p}) + 
		\sum\limits_{i \in \mathcal{R}^\text{N}/j} g_i(\mathbf{p}) \right\rbrace $,
$g(\mathbf{p}) = \sum\limits_{j \in \mathcal{R}^\text{N}} g_j(\mathbf{p})$,
$f_j(\mathbf{p}) = \log_2\left( \sum\limits_{i\in {\mathcal{R}^\text{N}}} p_i|h(i,j)|^2 + N^\text{N}_j \right)$,
and $g_j(\mathbf{p}) = 
	\log_2\left( \textstyle\sum\limits_{i\in {\mathcal{R}^\text{N}/j}} p_i|h(i,j)|^2 + N^\text{N}_j \right)$.

We use the Frank-and-Wold procedure in \citep[Algorithm 2]{dc_pwralloc} to solve problem (\ref{problem_NcPa_dc}).
For brevity, the detailed procedure is omitted here.

\section{Simulation Results}

Similar to \cite{CYang_CoD2D}, we consider a square hotspot area with the side length $100$ m.
There are totally $F = 200$ files in the system.
The memory of each user is able to cache $N_C = 10$ files and
the most popular $100$ files in $G = 10$ file groups are cached by the users.
The peak transmit power for each user is $23~\text{dBm}$. 
We allocate CDLs and NDLs with the same bandwidth $W_\text{C} = W_\text{N} = 10~\text{MHz}$.
The channel between any two users is modeled as $h = g\alpha$,
where $g = 37.6 + 36.8 \log_{10}(d) ~(\text{dB})$ is the path-loss, 
$d$ (m) is the distance between the two users,
and $\alpha$ is the Rayleigh fading factor.
The noise power at each user is $-90~\text{dBm}$.
The D2D radius for NDLs is set as $r = 30~\text{m}$. 

\begin{figure}
	\centering
	\includegraphics[scale=0.76]{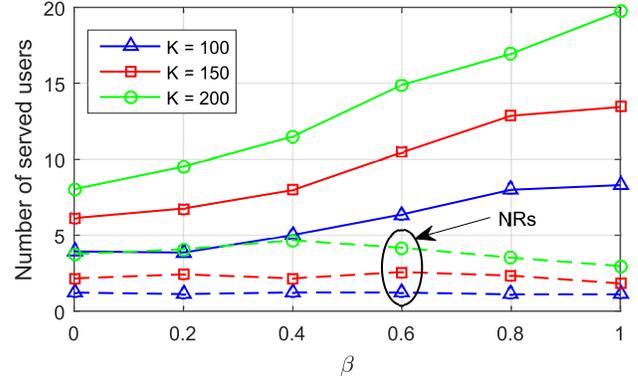}
	\caption{Average numbers of served users of CDLs and NDLs. }
	\label{fig_servedusers}
\end{figure}

In Fig. \ref{fig_servedusers}, the solid and dashed lines show respectively
the average numbers of scheduled CRs and NRs versus file popularity parameter $\beta$, with different numbers of users $K$.
The average number of scheduled CRs is larger than that of scheduled NRs,
which verifies the advantage of cooperation.
This is because CDLs are free from inter-link interference,
which improves the SINR at users and makes the QoS constraints of CDLs much easier to be satisfied
than that of NDLs.
Therefore, more CRs can be active.
The number of served CRs increases with $\beta$.
This is because with large $\beta$, the majority of users request a few popular files,
which leads to the increasing number of potential CRs,
since CDLs are established to deliver the mostly requested files.
On the other hand, the number of served NRs decreases with $\beta$ when $\beta$ is large.
This is because the number of potential NRs decreases with the increase of $\beta$.

\begin{figure}
	\centering
	\includegraphics[scale=0.76]{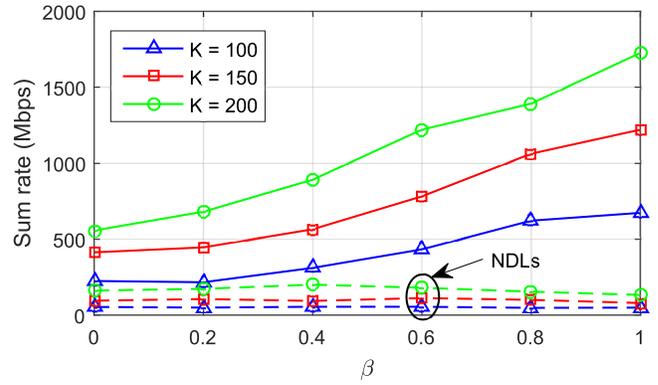}
	\caption{Average sum rates of CDLs and NDLs. }
	\label{fig_sumrate}
\end{figure}

In Fig. \ref{fig_sumrate}, the solid and dashed lines show respectively
the average sum rates of CDLs and NDLs versus $\beta$, with different $K$.
From the figure, CDLs have much higher sum rate than NDLs,
which also verifies the advantage of cooperation.
This is because CDLs are free from inter-link interference,
which increases the SINR and therefore the data rate of each user.
The sum rate of CDLs increases with $\beta$.
This is because more CRs are scheduled with larger $\beta$,
which coincides with Fig. \ref{fig_servedusers}.
On the other hand, the sum rate of NDLs decreases with $\beta$ when $\beta$ is large.
This is because fewer NDLs are scheduled with larger $\beta$,
which also coincides with Fig. \ref{fig_servedusers}.

\begin{figure}
	\centering
	\includegraphics[scale=0.76]{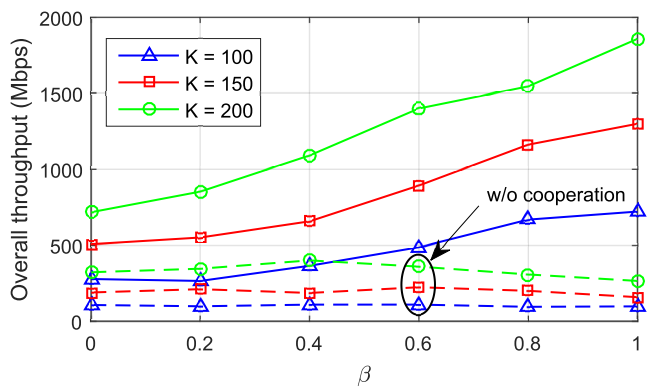}
	\caption{Average overall throughputs of D2D-enabled wireless caching networks with and without cooperation. }
	\label{fig_throughput}
\end{figure}

In Fig. \ref{fig_throughput}, the solid and dashed lines show respectively
the average overall throughputs of D2D-enabled wireless caching networks with and without cooperation versus $\beta$, with different $K$.
In the D2D-enabled wireless caching network without cooperation, we assume that all D2D links share the entire 
$W_\text{C} + W_\text{N}$ system bandwidth.
The figure shows that the overall throughput can be significantly improved with our proposed cooperative strategy.
The overall throughput with cooperation increases with $\beta$.
This is because more CDLs are established with higher $\beta$ and 
CDLs are free from inter-link interference and thus have much higher spectrum efficient than NDLs,
as we mentioned before.

\section{Conclusions}
In this paper, we study the resource allocation problem in
a cooperative D2D-enabled wireless caching network.
We formulate a joint link scheduling and power allocation problem to maximize the system throughput, 
which is NP-hard. 
To solve the problem, we decompose it into two sub-problems that optimize the CDLs and the NDLs, respectively. 
For CDLs, we propose a semi-orthogonal-based joint user scheduling and power allocation algorithm. 
For NDLs, we propose a novel low-complexity algorithm to perform link scheduling  
and a low-complexity D.C. programming method to solve the power allocation problem.
Simulation results show that the cooperative transmission can significantly improve both the number of served users
and the overall throughput of D2D-enabled caching networks.

\bibliographystyle{ieeetr}
\bibliography{references}
\end{document}